\newcommand{\option}{\iftrue}    % \iftrue  for arXiv
\documentclass{appolb_}
\usepackage{graphicx}
\usepackage{amsmath}
\usepackage{amssymb}

\usepackage{xcolor}
\definecolor{myblue}{rgb}{0.0, 0.0, 0.6}
\usepackage{hyperref}
\hypersetup{
  colorlinks = true,
  citecolor  = myblue,
  linkcolor  = myblue,
  urlcolor   = myblue
}
\begin{document}
% \eqsec  % uncomment this line to get equations numbered by (sec.num)
\title{Precision small scattering angle measurements of proton-proton and proton-nucleus analyzing powers at the RHIC hydrogen jet polarimeter%
\thanks{Presented at ``Diffraction and Low-$x$ 2022'', Corigliano Calabro (Italy), September 24--30, 2022.}%
}
\option
\author{A.~A.~Poblaguev, A.~Zelenski, E.~Aschenauer, G.~Atoian, K.~O.~Eyser, H.~Huang, W.~B.~Schmidke
  \address{Brookhaven National Laboratory,  Upton, New York 11973, USA}
  \\[3mm]
    {I.~Alekseev, D.~Svirida % of different affiliation
      \address{Alikhanov Institute for Theoretical and Experimental Physics,  117218, Moscow, Russia}
    }
    \\[3mm]
    N.~H.~Buttimore
    \address{School of Mathematics, Trinity College, Dublin 2, Ireland}
}
\else
\author{A.~A.~Poblaguev%
  ~for the RHIC Polarimetry Group\thanks{I.~Alekseev, E.~Aschenauer, G.~Atoian, N.~H.~Buttimore, K.~O.~Eyser, H.~Huang, A.~A.~Poblaguev, W.~B.~Schmidke, D.~Svirida, A.~Zelenski.}
  \address{Brookhaven National Laboratory,  Upton, New York 11973, USA}
  %, A.~Zelenski, I.~Alekseev, E.~Aschenauer, G.~Atoian, N.~H.~Buttimore, K.~O.~Eyser, H.~Huang, W.~B.~Schmidke, D.~Svirida
  %\address{RHIC Polarimetry Group}
}
\fi
\date{December 20, 2022}
\maketitle
\begin{abstract}
  At RHIC, the hydrogen jet target polarimeter (HJET) is used to measure proton beam polarization with accuracy $\sigma_P^\text{syst}/P\!\lesssim\!0.5\%$ by counting low energy (1--10\,MeV) recoil protons in left-right symmetric detectors. The HJET performance also allowed us to precisely measure $pp$ and $p$A (where A is any ion stored at RHIC) analyzing powers in the CNI region. The results of the measurements are discussed.
\end{abstract}
  
\section{Introduction}

The Polarized Atomic Hydrogen Gas Jet Target (HJET) \cite{Zelenski:2005mz} is employed to measure absolute vertical polarization of the high energy ($\sim\!100\,\text{GeV}$) proton beams at the Relativistic Heavy Ion Collider (RHIC). The jet polarization is about $P_j\!\sim\!96\%$ and is monitored with accuracy of about $0.1\%$. 

\begin{figure}[t]
  \begin{minipage}[t]{0.55\columnwidth}
    \begin{center}
      \includegraphics[width=1.0 \columnwidth]{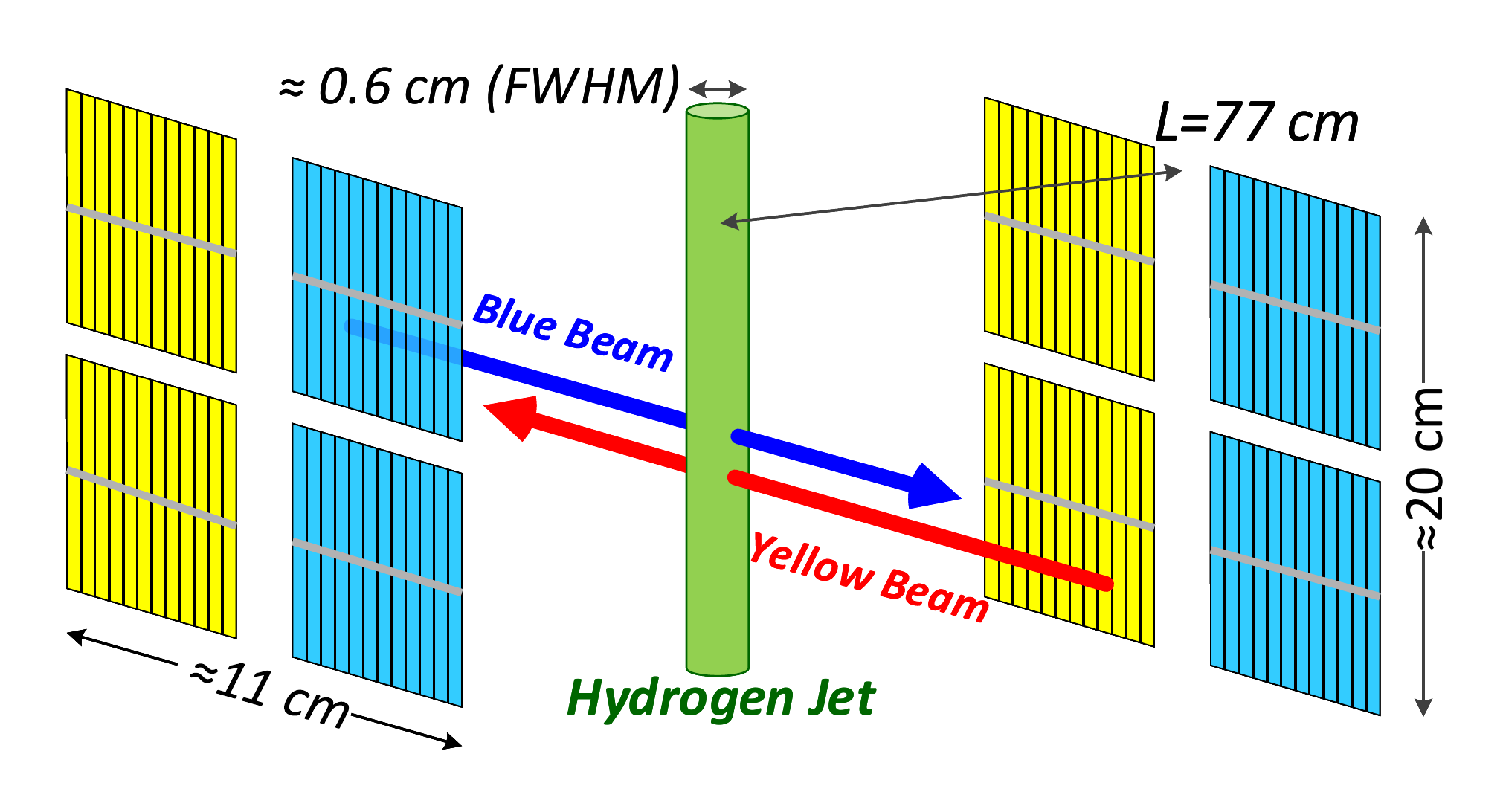}
    \end{center}
    \caption{ A schematic view of the HJET recoil spectrometer.
    }
    \label{fig:HjetView}
  \end{minipage}
  \hfill
  \begin{minipage}[t]{0.43\columnwidth}
    \begin{center}
      \includegraphics[width=0.9\columnwidth]{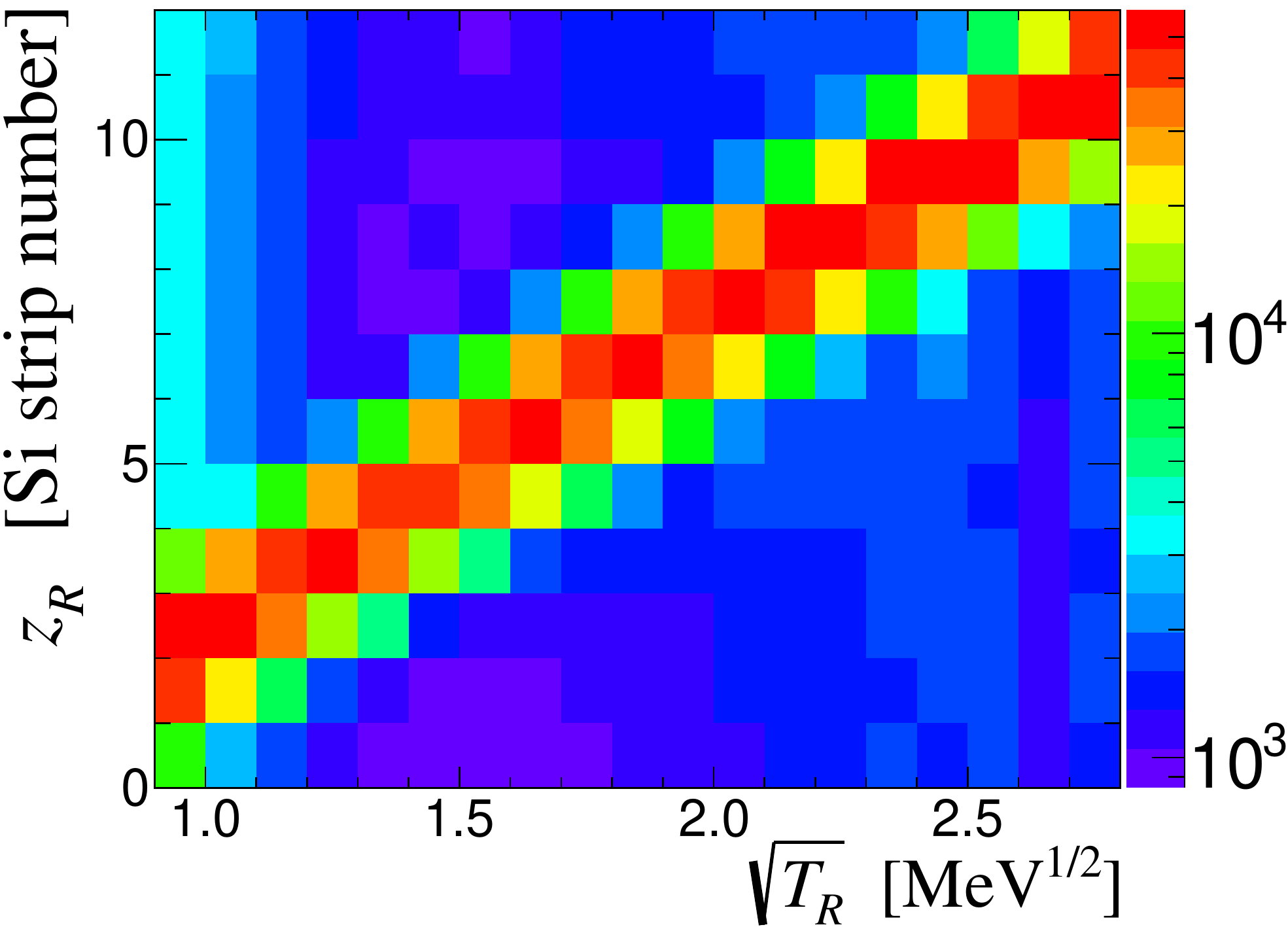}
    \end{center}
    \caption{
      Measured $z_R$ {\em vs} $\sqrt{T_R}$ for\,the 100\,GeV proton beam.}
    \label{fig:zR-TR}
  \end{minipage}
\end{figure}

The recoil protons from the RHIC beam scattering off the jet are counted in the left-right symmetric Si strip detectors depicted in Fig.\,\ref{fig:HjetView}. For each proton detected, time of flight (ToF), kinetic energy $T_R$, and coordinate~$z_R$ (along the beam) in the detector discriminated by the strip width of 3.75\,mm are determined. Detailed description of the measurements is given in Ref.\,\cite{Poblaguev:2020qbw}.

The following kinematical relations are important for the data analysis
\begin{equation}
  t = -2m_pT_R,
\end{equation}
\begin{equation}
  \frac{z_R-z_\text{jet}}{L}=%
  \sqrt{\frac{T_R}{2m_p}}\times
  \left[ 1+\frac{m_p}{E_\text{beam}}%
    \left(\frac{m_p}{M}+\frac{\Delta}{T_R}\right)%
    \right],
\end{equation}
where $t$ is the momentum transfer squared, $m_p$ is the proton mass, $E_\text{beam}$ is the beam energy per nucleon, $M$ is the beam particle mass, $\Delta\!=\!M_X\!-\!M$ with $M_X$ being the effective scattered mass, and $z_\text{jet}$ ($\langle{z_\text{jet}}\rangle\!=\!0$, $\langle{z_\text{jet}^2}\rangle^{1/2}\!\approx\!2.5\,\text{mm}$) is the coordinate of the scattering point.

The HJET detector
geometry allows us to make measurements only in the Coulomb-nuclear interference (CNI) low momentum transfer range  $0.0013\!<\!-t\!<\!0.018\,\text{GeV}^2$ ($0.6\!<\!T_R\!<\!10\,\text{MeV})$, which is nearly independent of the beam particle mass and energy. 

For recoil protons, measured $\text{ToF}$ and $T_R$ must be kinematically consistent. Considering the correlation shown in Fig.\,\ref{fig:zR-TR}, one can easily identify the elastic events ($\Delta\!=\!0$). The background rate (as a function of $T_R$) can be interpolated\,\cite{Poblaguev:2020qbw}, with a relative accuracy of about 3\%, to the elastic values of $z_R(T_R)$. Thus, the background can be accurately subtracted from the elastic data, which allows a low, $\sigma_P^\text{syst}/P\!\lesssim\!0.5\%$, systematic uncertainty in the beam polarization measurement. 

\section{Spin asymmetries measured with the HJET}

\begin{figure}[b]
  \begin{minipage}[t]{0.65\columnwidth}
    \begin{center}
      \includegraphics[width=0.49\columnwidth]{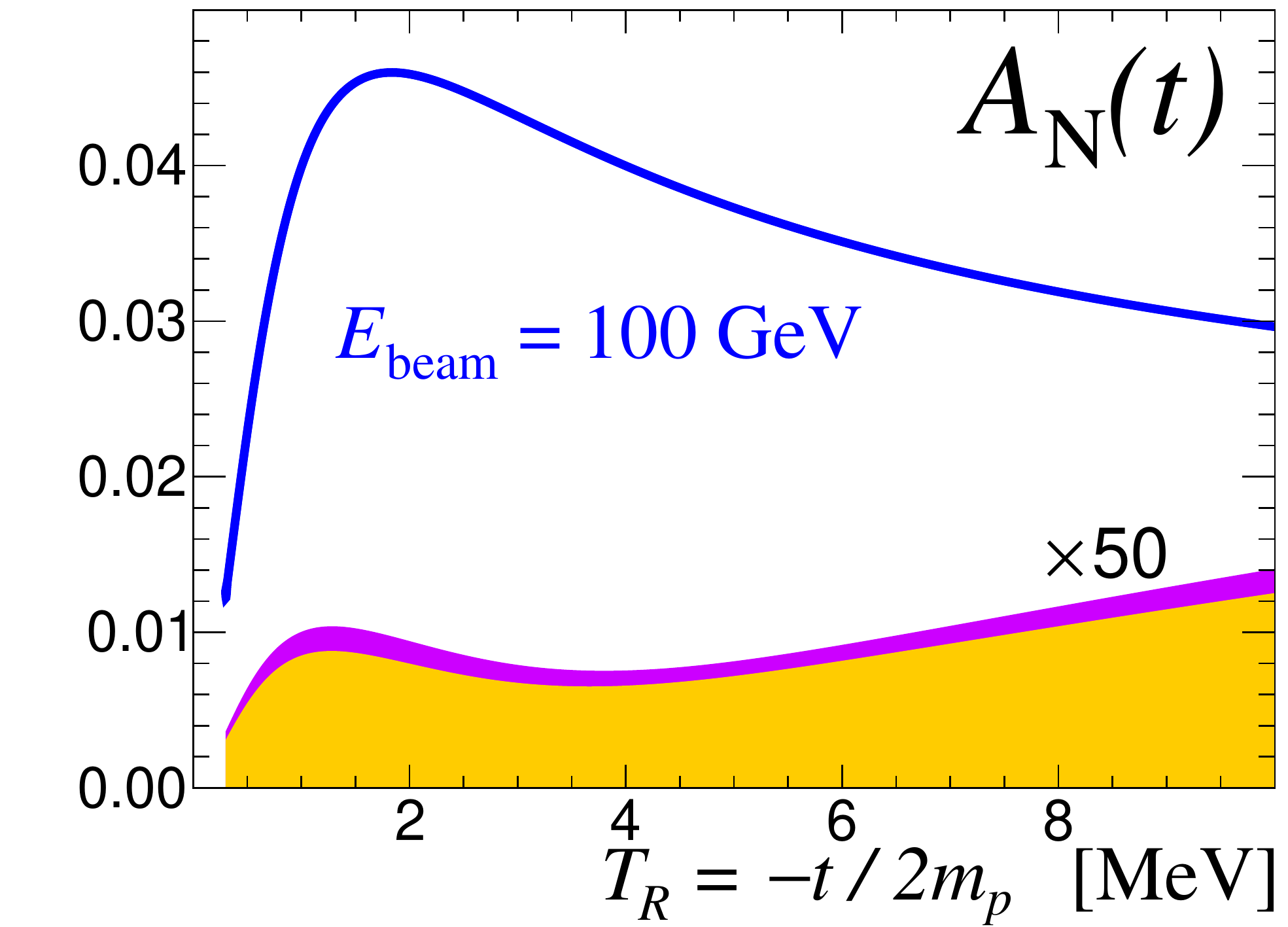}
      \hfill
      \includegraphics[width=0.49\columnwidth]{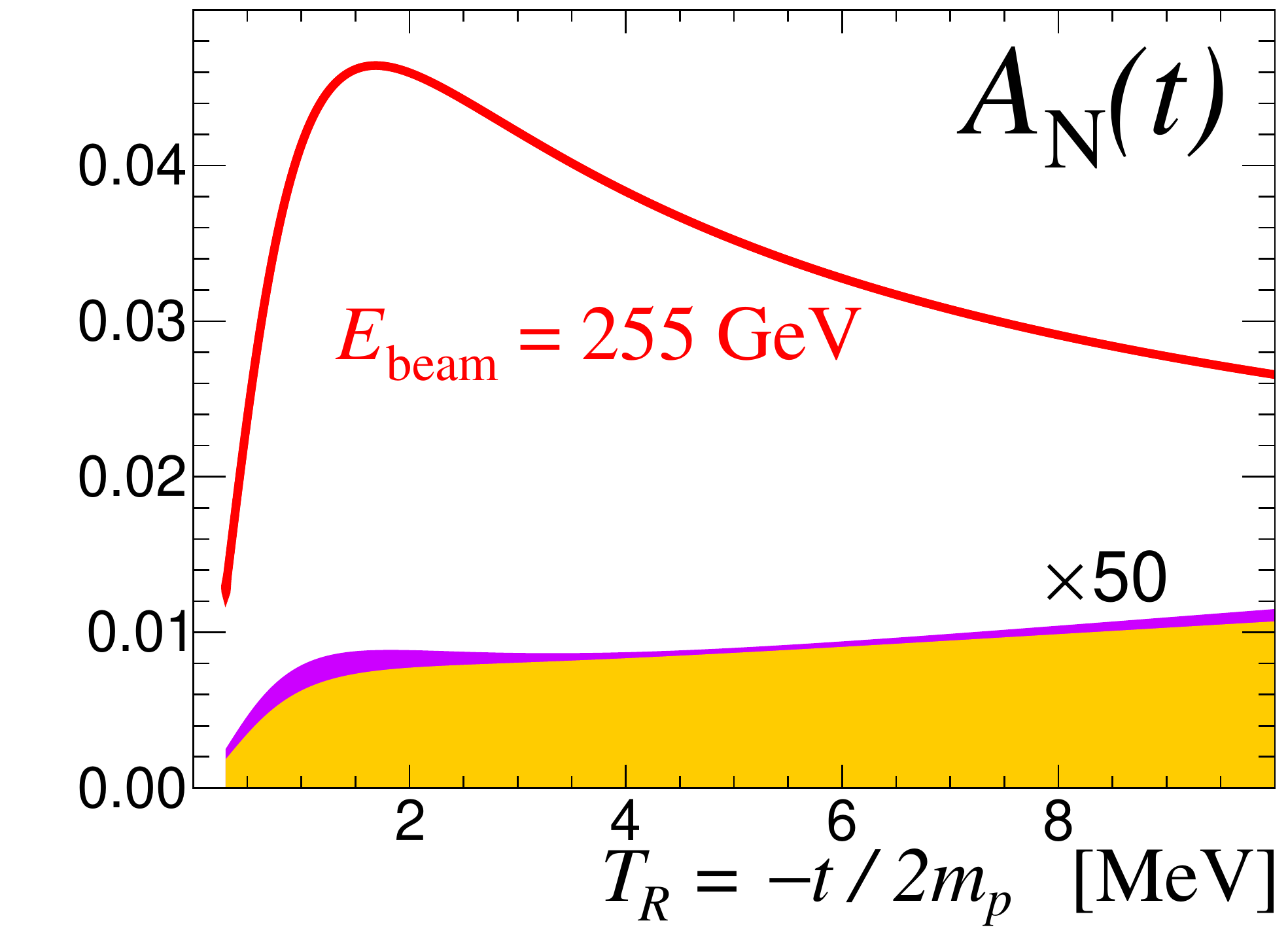}
    \end{center}
    \caption{Single spin-flip elastic $pp$ $A_\text{N}(t)$. Filled areas show experimental uncertainties scaled by a factor of 50. Violet is for {\em stat+syst} and orange is for {\em syst} only.
    }
    \label{fig:AN}
  \end{minipage}
  \hfill
  \begin{minipage}[t]{0.325\columnwidth}
    \begin{center}
      \includegraphics[width=0.98\columnwidth]{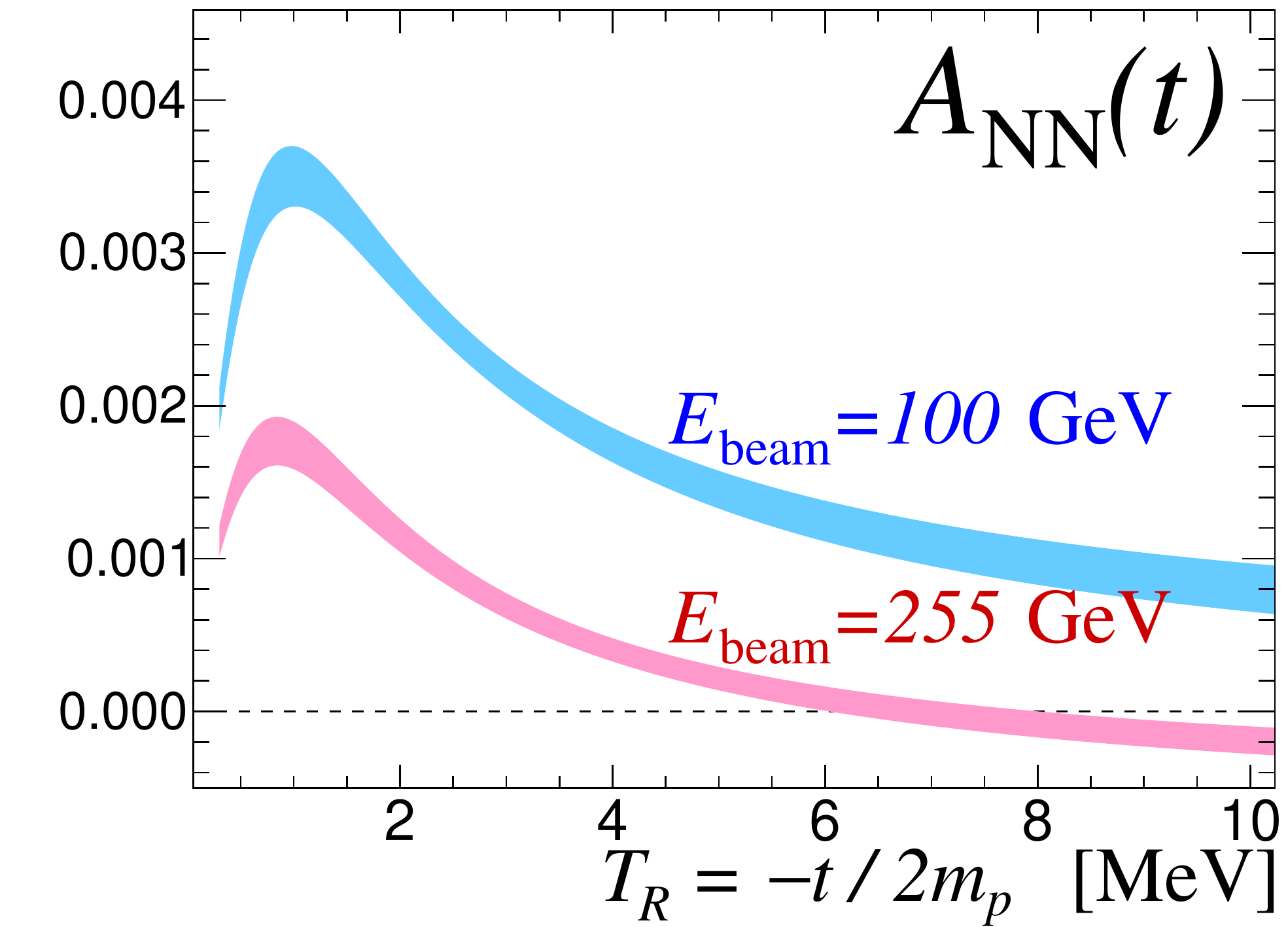}
    \end{center}
    \caption{Double spin-flip forward elastic $pp$ analyzing power $A_\text{NN}(t)$.
    }
    \label{fig:ANN}
  \end{minipage}
\end{figure}

For vertically polarized beam ($b$) and target ($j$), the recoil proton azimuthal angle distribution can be written as
\begin{equation}
\frac{d^2\sigma}{dtd\varphi}\!=\!\frac{d\sigma}{2\pi dt}\,%
\Big[1+\left(A_\text{N}^jP_j\!+\!A_\text{N}^bP_b\right)\!\sin\;\!\!{\varphi}% 
  +\left(A_\text{NN}\sin^2\:\!\!\!{\varphi}\!+\!A_\text{SS}\cos^2\:\!\!\!{\varphi}\right)\!P_bP_j  \Big].
\label{eq:dsdt}
\end{equation}
Since $\cos{\varphi}\!\approx\!0$ at HJET, the measurements are insensitive to $A_\text{SS}(t)$. For elastic
polarized $p^\uparrow p^\uparrow$
%proton-proton
scattering, $A_N^b=A_N^j=A_N(t)$ and the asymmetries, $a_\text{N}^{b,j}(T_R)=A_\text{N}^{b,j}(t)P_{b,j}$, $a_\text{NN}(T_R)=A_\text{NN}(t)P_bP_j$ concurrently measured\,\cite{Poblaguev:2020qbw} using the same events allow one to determine $P_b$, $A_N(t)$, and $A_\text{NN}(t)$.

Theoretical parametrization of the forward elastic proton-proton analyzing powers was developed in Refs.\,\cite{Kopeliovich:1974ee,Buttimore:1978ry,Buttimore:1998rj}. In a simplified form,
\begin{equation}
  A_\text{N}(t) = \frac%
  {\sqrt{-t}}{m_p}\,\frac{(\kappa_p-2\text{Im}\,r_5)\,t_c/t - 2\text{Re}\,r_5}%
  {(t_c/t)^2-2(\rho+\delta_C)\,t_c/t+1},
\end{equation}
where $\kappa_p\!=\!1.792$ is the anomalous magnetic moment of a proton, $t_c\!=\!-8\pi\alpha/\sigma_\text{tot}$, $\sigma_\text{tot}$ is the total $pp$ cross section, $\rho$ is the Re/Im amplitude ratio, $\delta_C\!\approx\!\alpha\ln{t_c/t}\!+\!0.024$ is the Coulomb phase, and $|r_5|\!\sim\!0.02$ is the hadronic single spin-flip amplitude parameter.
A more accurate expression discussed in \cite{Poblaguev:2019vho,Poblaguev:2021xkd} includes small but essential corrections to meet the experimental precision achieved at HJET.

\section{Forward elastic proton-proton analyzing powers}

During the RHIC Runs in 2015 and 2017 with beam energies of 100 and 255\,GeV, respectively, the forward elastic proton-proton $A_\text{N}(t)$ (Fig.\,\ref{fig:AN}) and $A_\text{NN}(t)$ (Fig.\,\ref{fig:ANN}) were precisely measured\,\cite{Poblaguev:2019saw} at HJET. 

\begin{figure}[t]
  \option
  \begin{minipage}[c]{0.34\columnwidth}
  \else
  \begin{minipage}[c]{0.51\columnwidth}
  \fi
  \begin{center}
    \includegraphics[width=1.\columnwidth]{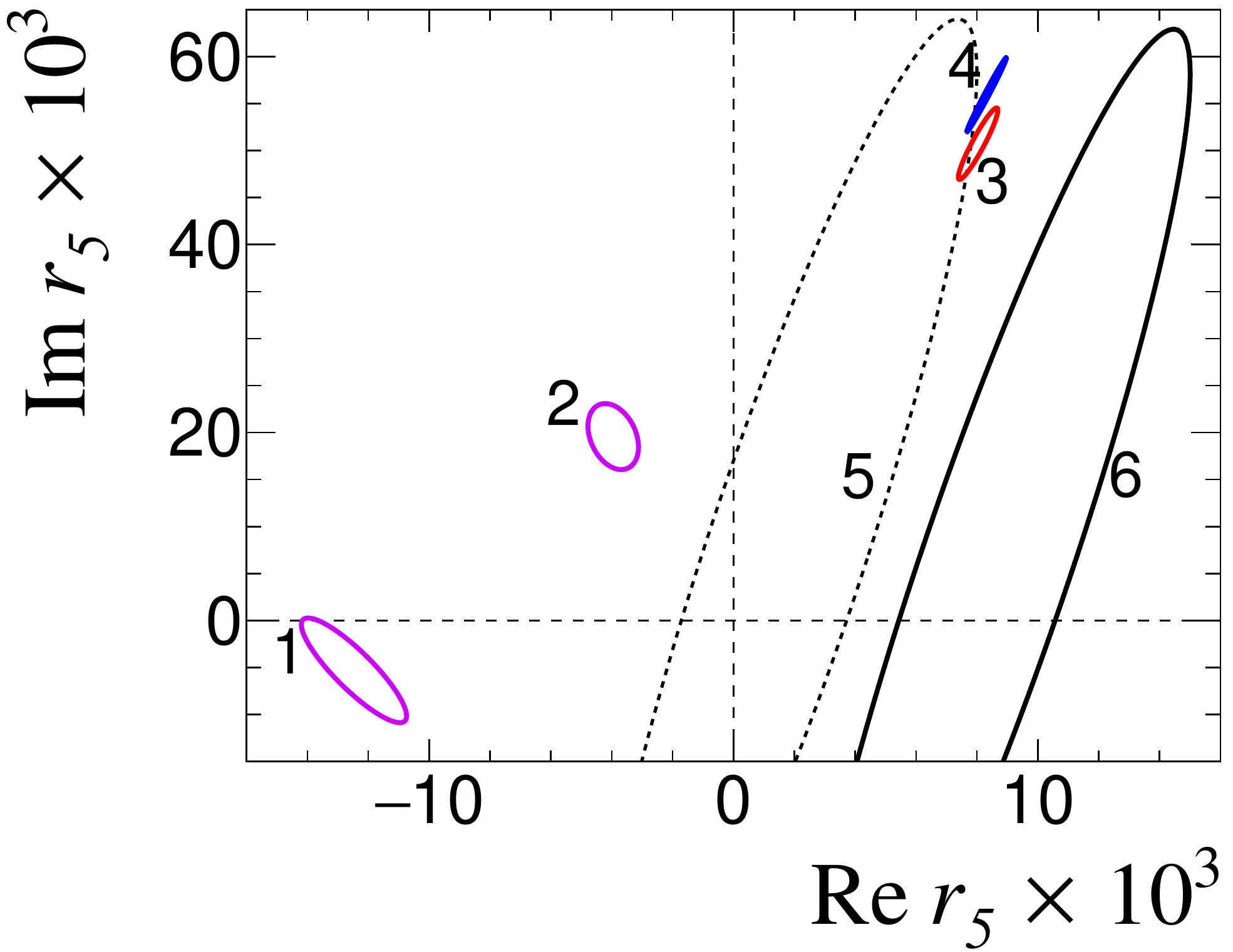}
  \end{center}
  \end{minipage}
  \hfill
  \option
  \begin{minipage}[c]{0.64\columnwidth}
  \else
  \begin{minipage}[c]{0.46\columnwidth}
  \fi
  \caption{Experimental 1-sigma contours (stat+syst) for $r_5$. The HJET results are marked \lq\lq1\rq\rq\ ($\sqrt{s}\!=\!13.76$ GeV) and \lq\lq2\rq\rq\ (21.92\,GeV). The Regge fit extrapolations to 200\,GeV are \lq\lq3\rq\rq\ for Froissaron and \lq\lq4\rq\rq\ for simple pole Pomeron. \lq\lq5\rq\rq\ and \lq\lq6\rq\rq\:are the STAR value (200\,GeV) \cite{STAR:2012fiw} before and after applying (in this analysis) the absorption and $r_p$ related corrections.
    }
  \label{fig:Regge_r5}
  \end{minipage}
\end{figure}

The hadronic single spin-flip amplitude ($r_5$), was clearly isolated as shown in Fig.\,\ref{fig:Regge_r5}. Here, compared to the published\,\cite{Poblaguev:2019saw} results, we applied a correction $\text{Re}\,r_5=\text{Re}\,r_5^{\cite{Poblaguev:2019saw}}+(3.1_\text{abs}+0.8_{r_p})\times10^{-3}$ due to the absorption\,\cite{Poblaguev:2021xkd} and updated value of the proton charge radius $r_p\!=\!0.841\,\text{fm}$\,\cite{Workman:2022ynf}.

To find the spin-flip amplitude dependence on the center of mass energy squared, $s=2m_p(m_b+E_\text{beam})$, the following parametrization was used
\begin{equation}
  \sigma_\text{tot}(s)\,r_5(s) = f_5^+R^+(s)+f_5^-R^-(s)+f_5^PP(s),
  \label{eq:r5Fit}
\end{equation}
where couplings $f_5^{\pm,P}$ are free parameters in the fit and Reggeon pole $R^\pm(s)$ and Pomeron $P(s)$ (in Froissaron approximation) are functions found in a fit\,\cite{Fagundes:2017iwb} $\sigma_\text{tot}(s)\left[i+\rho(s)\right]\!=\!R^+(s)\!+\!R^-(s)\!+\!P(s)$ of the unpolarized $pp$ data.

The $r_5$ fit result, $f_5^P=0.054\pm0.002_\text{stat}\pm0.003_\text{syst}$, suggests non-vanishing hadronic spin-flip amplitude at very high energies. The extrapolation of the HJET values of $r_5$ to $\sqrt{s}=200\,\text{GeV}$ can be compared with the STAR measurement\,\,\cite{STAR:2012fiw}.

The correction applied to $r_5$ improves the agreement of the HJET results with the parametrisation in Eq.\,(\ref{eq:r5Fit}), since $\chi^2\!=\!2.2\!\to\!0.7$\,(ndf=1). Although consistency with STAR value of $r_5$ dis-improves, the discrepancy is not critically significant, $\chi^2/\text{ndf}\!=\!4.8/3$ (which is statistically equivalent to about 1.8 standard deviations).

For the $A_\text{NN}(t)$, the hadronic double spin-flip amplitude ($r_2$) is also non-zero and the Regge fit gives $f_2^P\!=\!-0.0020\!\pm\!0.0002$ \cite{Poblaguev:2019saw}.

\begin{figure}[t]
  \begin{center}
    \includegraphics[width=0.32\columnwidth]{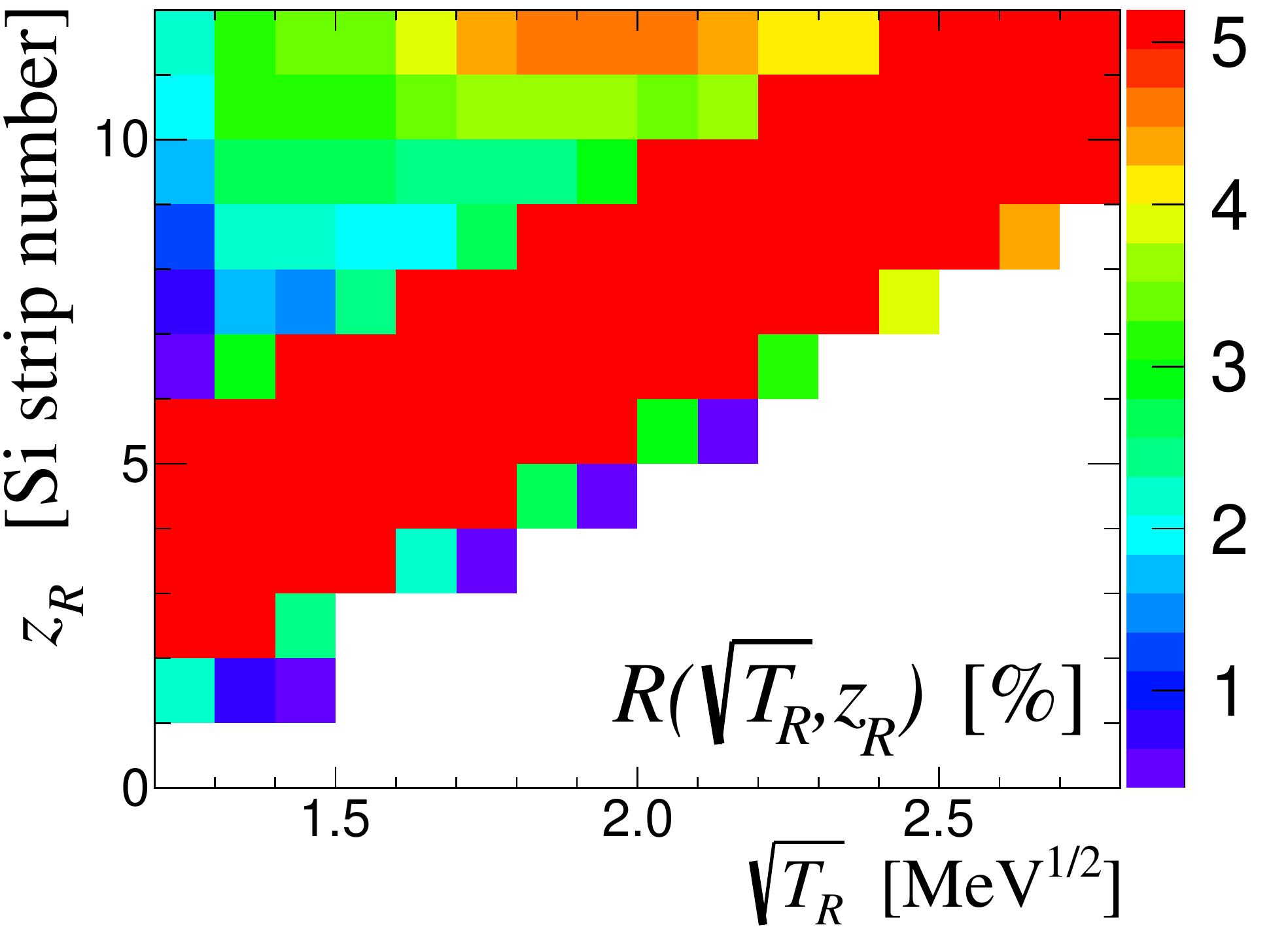}
    \includegraphics[width=0.32\columnwidth]{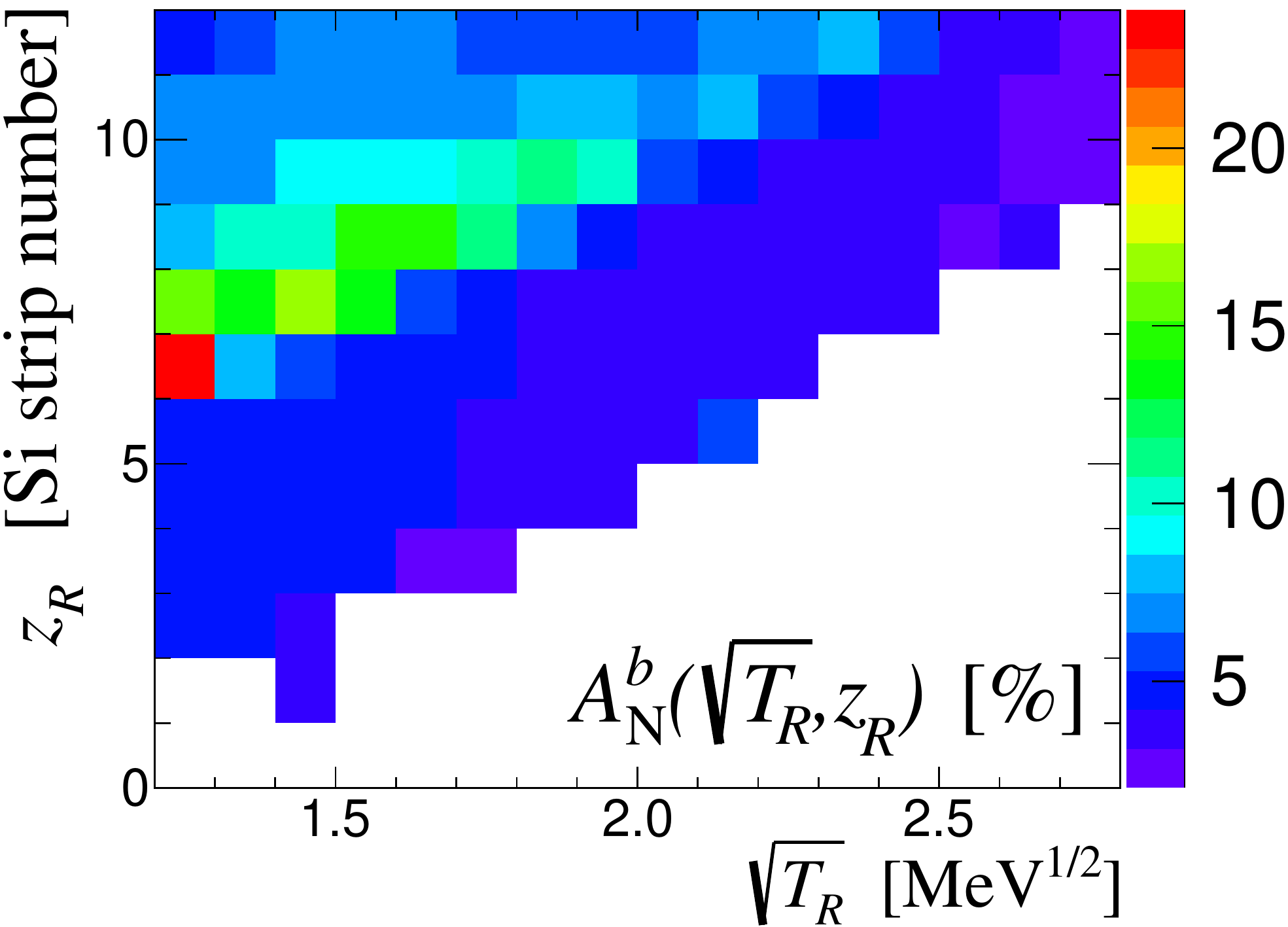}
    \includegraphics[width=0.32\columnwidth]{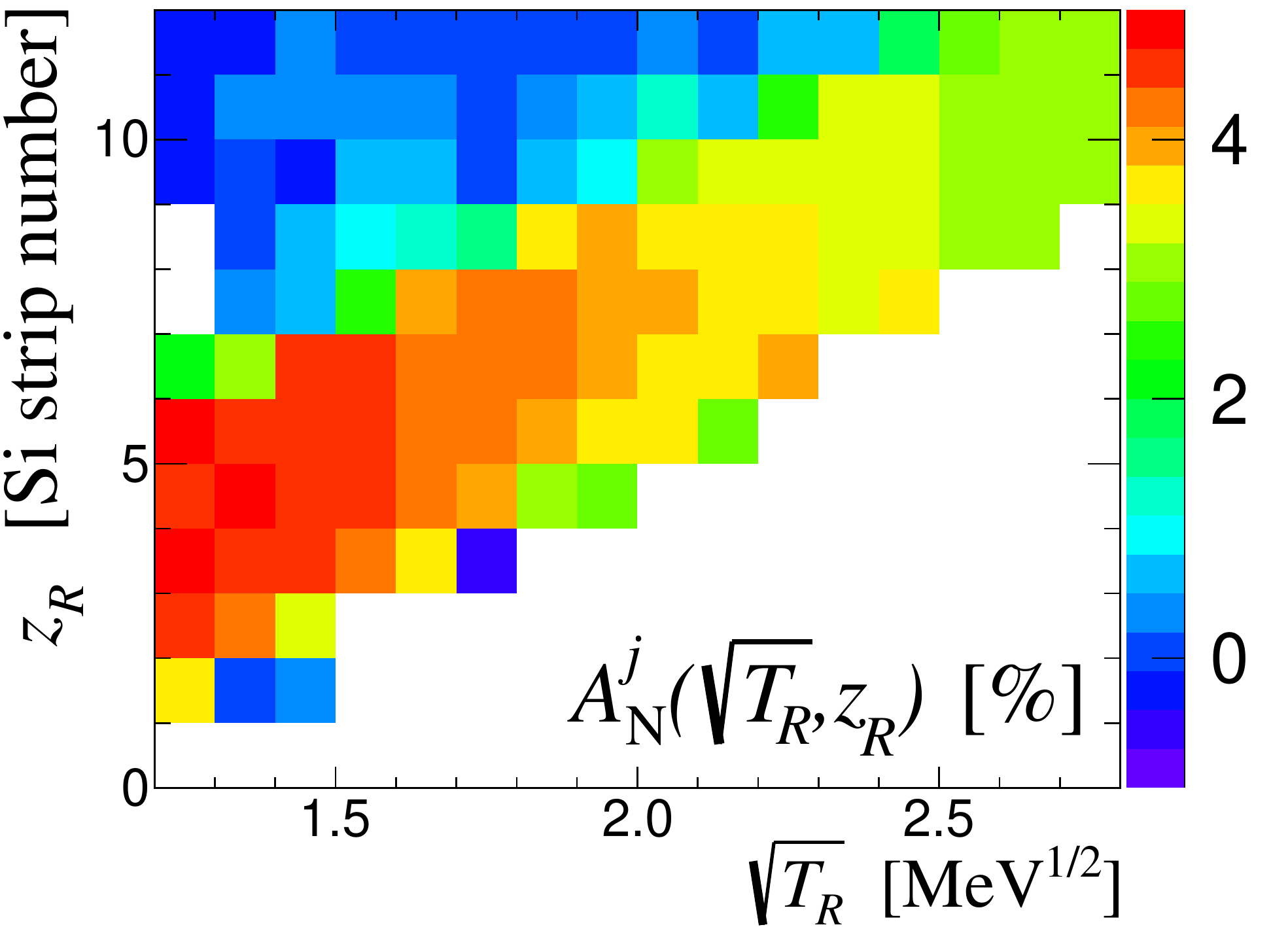}
  \end{center}
  \caption{Shown are elastic and inelastic events for the 255\,GeV proton
beam, left to right, normalized event rate $R(\sqrt{T_R},z_R)\!=\!N_\text{bin}(\sqrt{T_R},z_R)/N_\text{max}^\text{el}$, and the beam $A_\text{N}^b(t,\Delta)$ and target $A_\text{N}^j(t,\Delta)$ analyzing powers. A cutoff of $R>0.4\%$ was used.
  }
  \label{fig:ppInel255}
\end{figure}

\begin{figure}[t]
  \begin{center}
    \includegraphics[width=0.32\columnwidth]{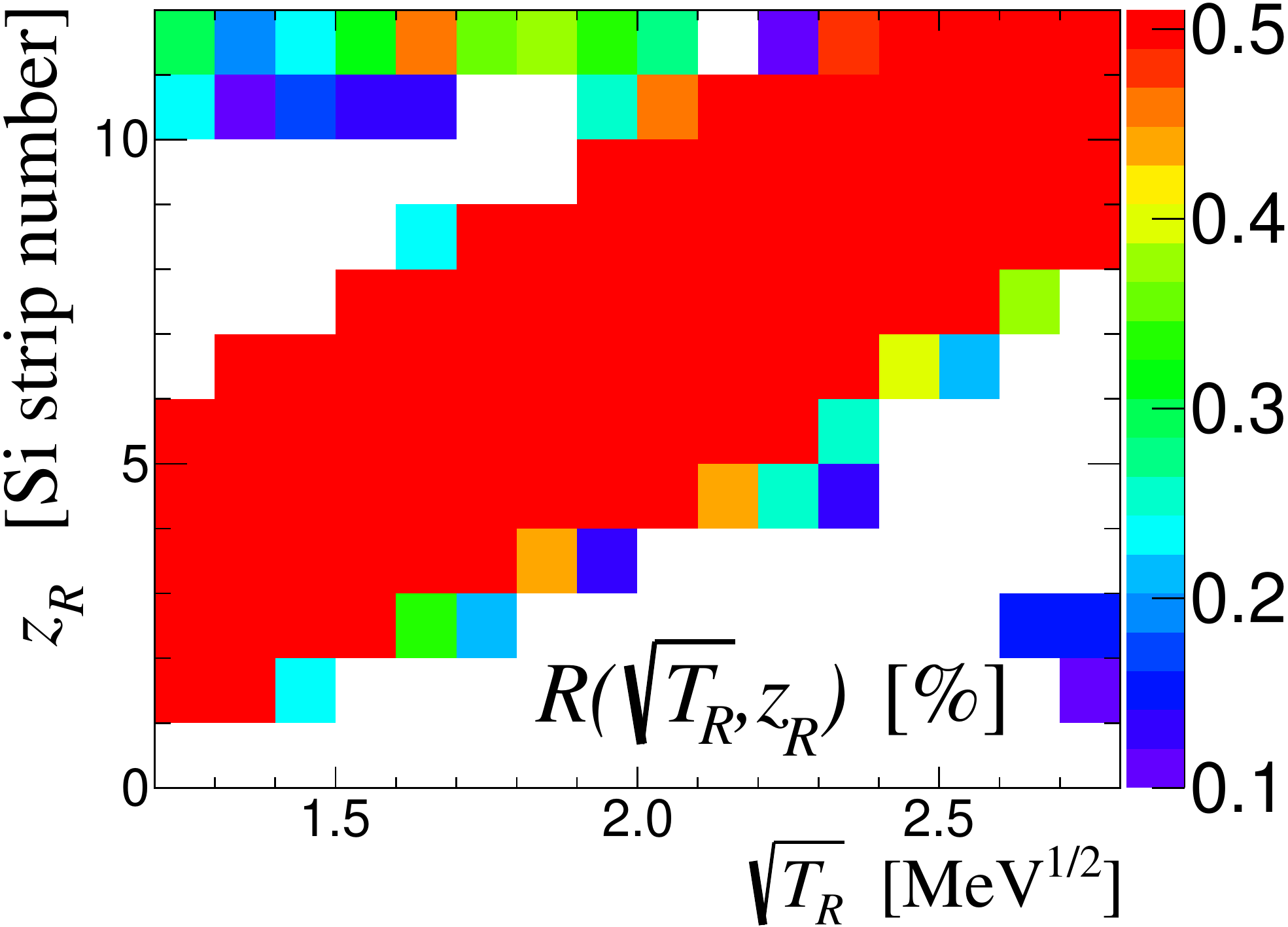}
    \includegraphics[width=0.32\columnwidth]{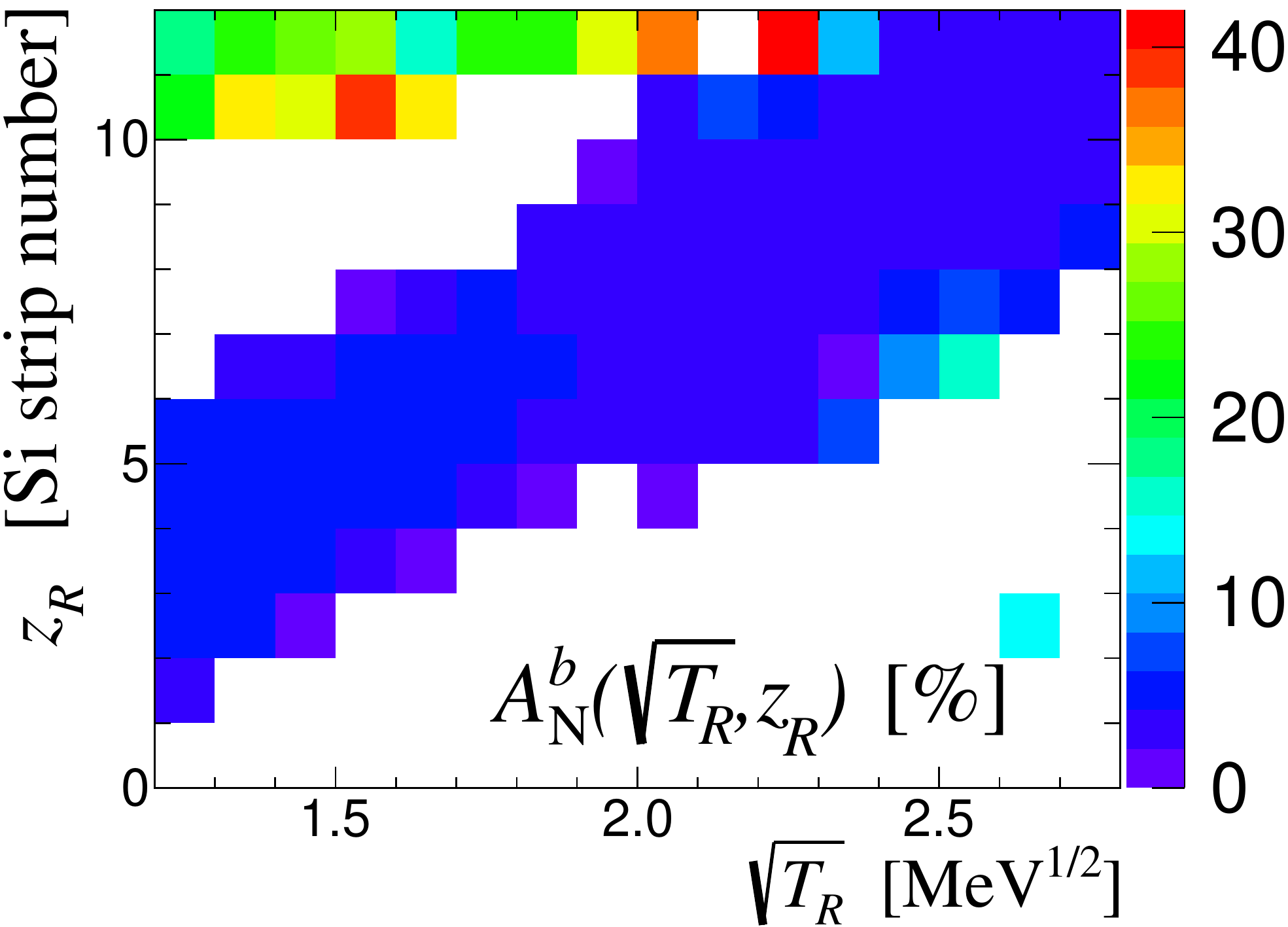}
    \includegraphics[width=0.32\columnwidth]{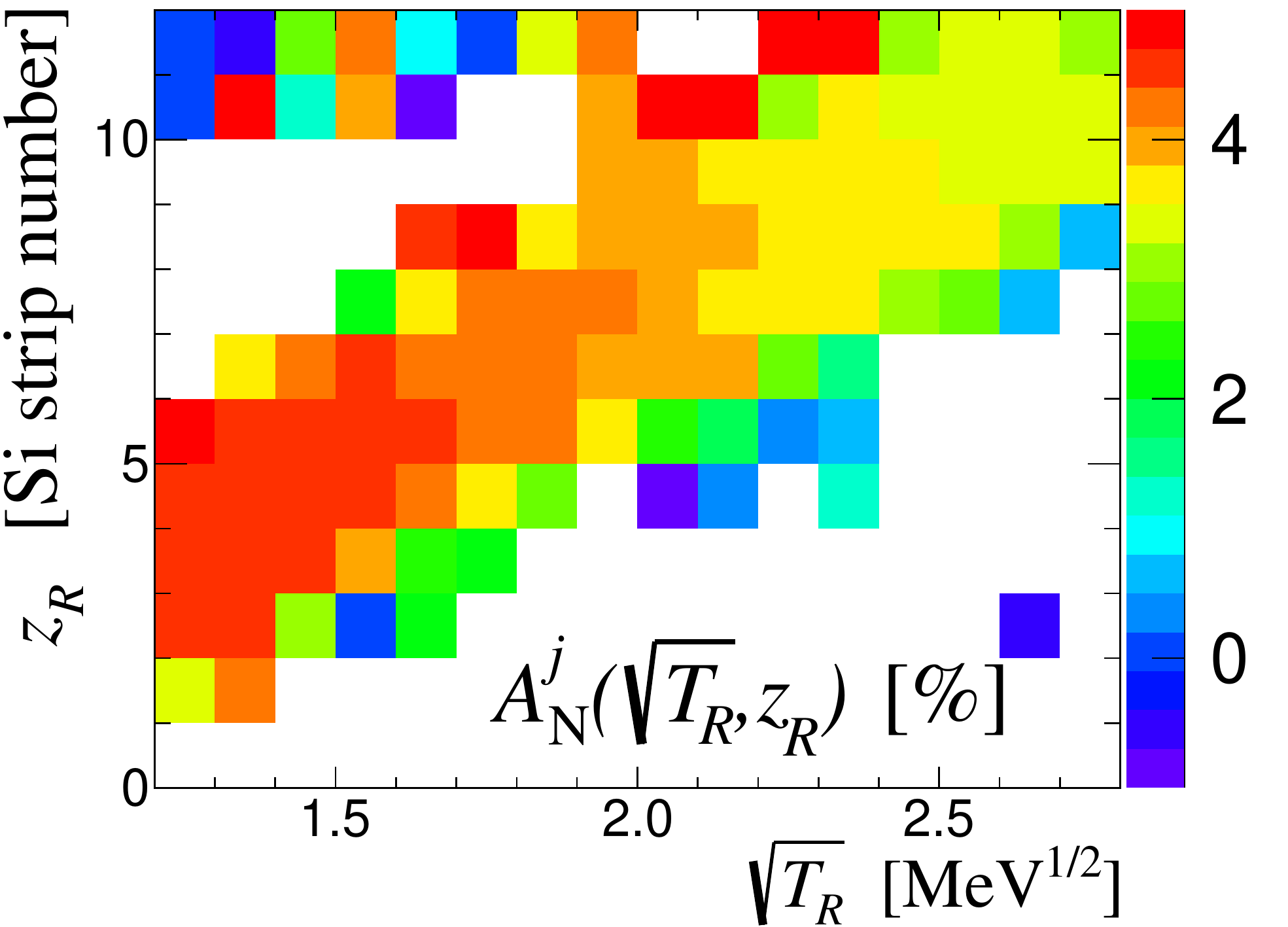}
  \end{center}
  \caption{Same as Fig.\,\ref{fig:ppInel255}, but for the 100\,GeV proton beam and $R>0.1\%$ cutoff.}
  \label{fig:ppInel100}
\end{figure}

\section{Inelastic proton-proton scattering}

With the HJET detectors, the inelastic events $p_b\,p_j\!\to\!(\pi X)_b\,p_j$ can be separated (due to relatively large $\Delta\!\ge\!m_\pi$) from the elastic ones and the inelastic analyzing powers for the beam $A_\text{N}^b(t,\Delta)$ and target $A_\text{N}^j(t,\Delta)$ spins can be evaluated.

Preliminary results for the 255\,GeV proton beam are shown in Fig.\,\ref{fig:ppInel255}. Only bins with event rate (after background subtraction) $R>0.4\%$ relative to the elastic maximum were analyzed. The inelastic events ($R$ up to 5\%) are well identified in the upper left corner of the histograms.

One can see that $A_\text{N}^j\!<\!A_\text{N}^\text{elastic}\!<\!A_\text{N}^b$. The inelastic analyzing power grows with decreasing of $\Delta$. For $A_\text{N}^b(t,\Delta)$, values of about 20\% are observed in the data.

For the 100\,GeV beam (Fig.\,\ref{fig:ppInel100}), the detected inelastic rate is much lower, $R\lesssim0.5\%$. Nonetheless, results for the analyzing powers are, qualitatively, about the same as for 255\,GeV. A 100\,GeV beam spin inelastic analyzing power up to 35\% was observed.

\section{Elastic proton-nucleus analyzing power}

\begin{figure}[t]
  \begin{center}
    \includegraphics[width=0.48\columnwidth]{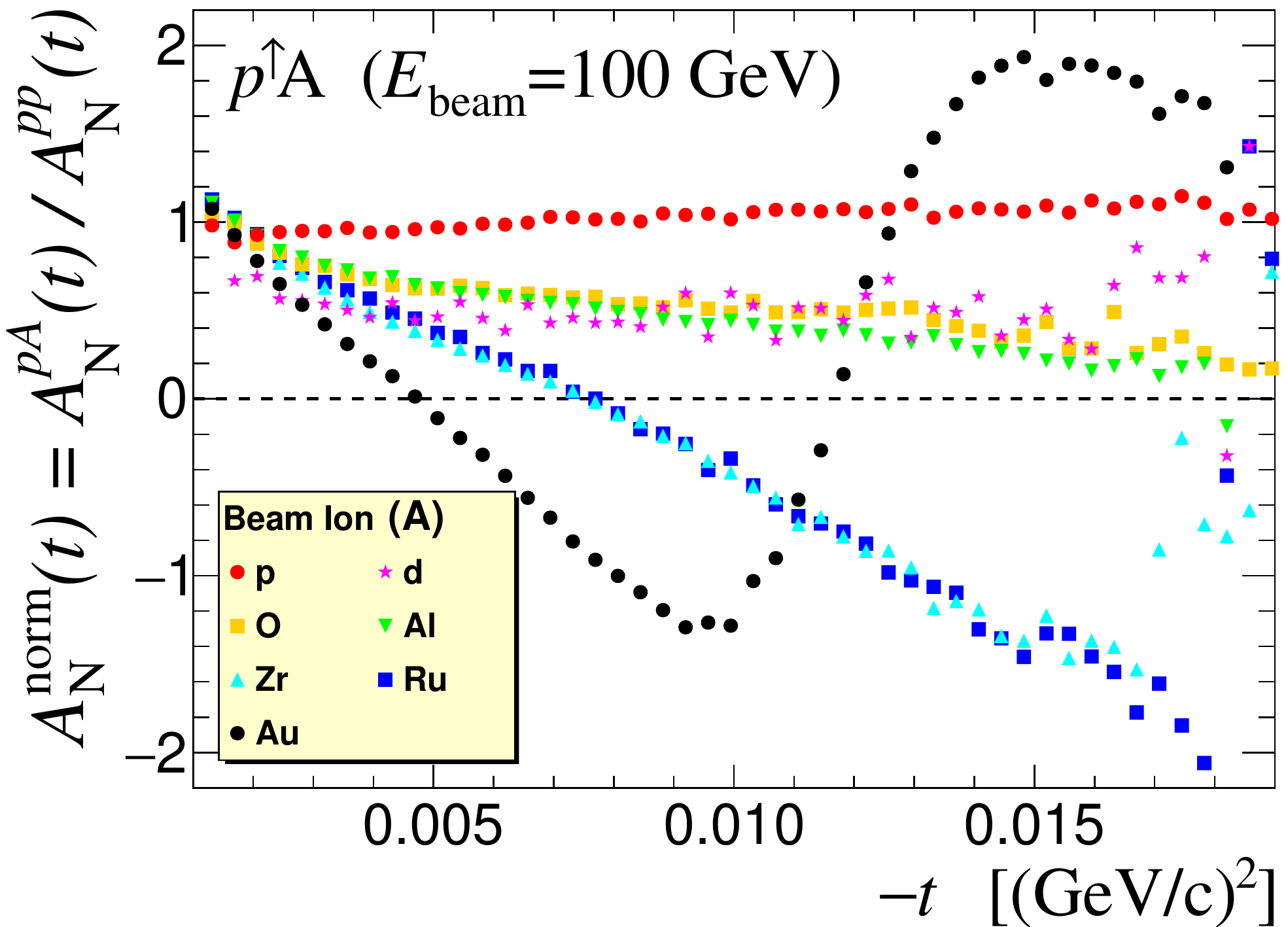}
    \includegraphics[width=0.48\columnwidth]{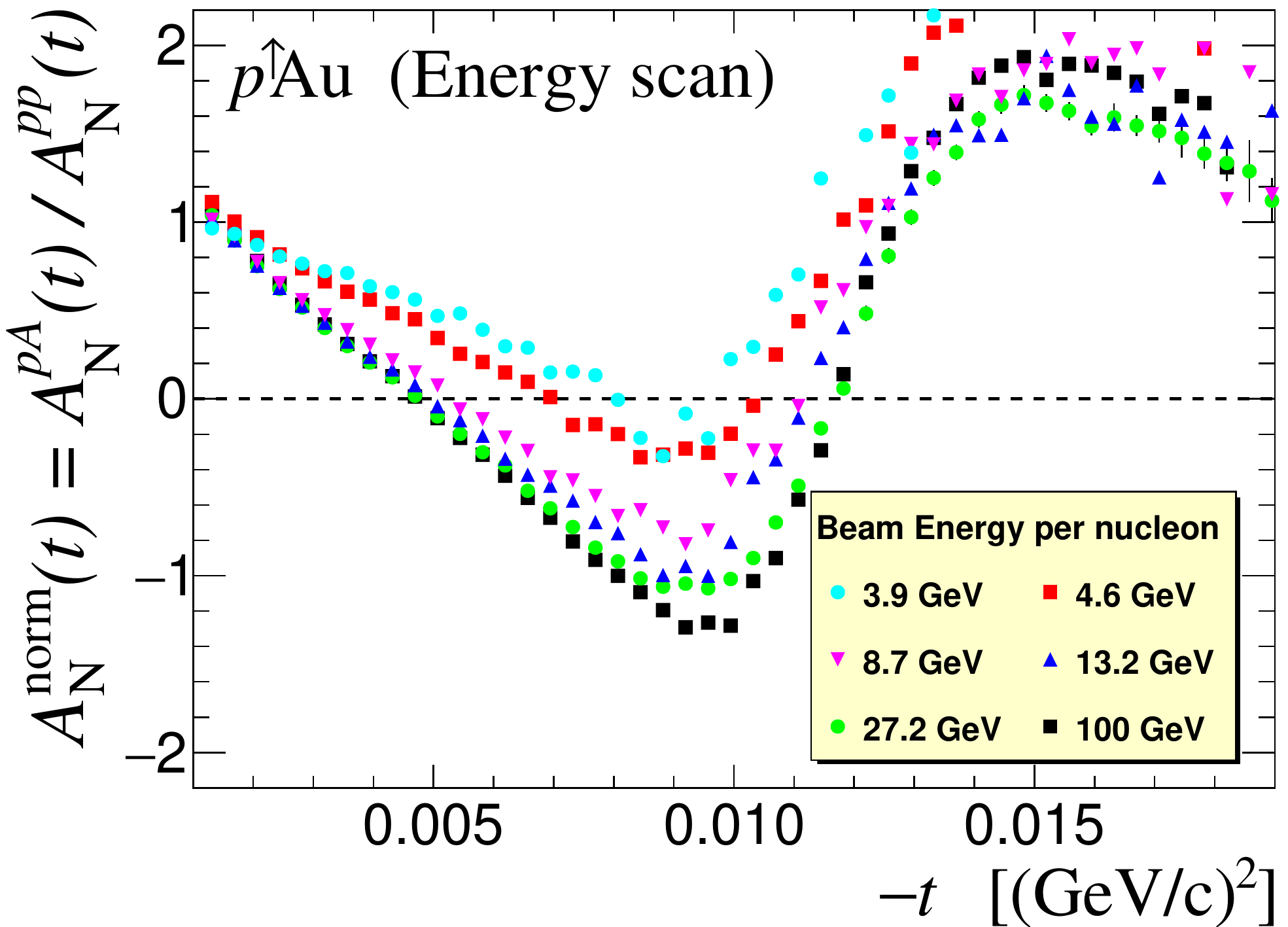}
  \end{center}
  \caption{
    The dependence of the proton-nucleus elastic $A_\text{N}^{pA}(t)$ on
the beam ion (left) and the beam energy (right). The measured
analyzing powers are normalized by the
${pp}$ one calculated for $E_\text{beam}\!=\!100\,\text{GeV}$ and no hadronic single spin-flip, $r_5\!=\!0$.
  }
  \label{fig:pA}
\end{figure}

Since 2015, HJET was routinely operated in RHIC Heavy Ion Runs. It was found that HJET performance in an ion beam is as good as in a proton one. Consequently, the proton-nucleus analyzing power $A_\text{N}^{pA}(t)$ can be precisely measured. The study has been performed for six ions [${}^2$H\,($d$),\,${}^{16}$O, ${}^{27}$Al,\,${}^{96}$Zr,\,${}^{96}$Ru,\,and\,${}^{197}$Au]. Also, the energy scans were done for Au and $d$.

Some preliminary results for normalized, $A_\text{N}^\text{norm}(t)\!=\!A_\text{N}^{pA}(t)/A_\text{N}^{pp}(t)$, analyzing powers are shown in Fig.\,\ref{fig:pA}. Systematic uncertainties in the measurements were not considered. The beam nucleus breakup fraction in the elastic data is expected to be small, $\lesssim\!1\%$\,\cite{Poblaguev:2022hsi}.

For 100\,GeV/nucleon Au beam, the experimental data were compared with theoretical predictions in Ref.\,\cite{Krelina:2019mlu}. It was found that absorption corrections are very important for calculation of $A_\text{N}^{p\text{Au}}(t)$. However, not all essential discrepancies between the data and theory were eliminated. 

\section{Summary}

The HJET, which was designed to measure absolute proton beam polarization at RHIC, can also be considered as a standalone fixed target experiment to precisely measure proton-proton and proton-nucleus analyzing powers in the CNI region.

The measurements of elastic $pp$ $A_\text{N}(t)$ and $A_\text{NN}(t)$ resulted in finding non-zero single and double spin Pomeron couplings.

Preliminary results for inelastic $pp$ $A_\text{N}^b(t,\Delta)$ and $A_\text{N}^j(t,\Delta)$ for 100 and 255\,GeV, and elastic $p\text{A}$ analyzing powers, in wide range of $1\!<\!\text{A}\!<\!200$ (for $E_\text{beam}\!=\!100\,\text{GeV}$) and $3.8\!<E_\text{beam}\!<\!100\,\text{GeV}$ (for Au), were obtained. However, to properly understand these measurements, an appropriate theoretical description of these analyzing powers is needed.

\iffalse
\bibliographystyle{IEEEtran}                                       
\bibliography{../aapBib}
\else

\fi

\end{document}